# Uni-traveling-carrier photodetector with high-contrast grating focusing-reflection mirrors

Qingtao Chen[1,3], Wenjing Fang[2], Yongqing Huang[1*], Xiaofeng Duan[1], Kai Liu[1], Mohammad S. Sharawi[3], and Xiaomin Ren[1]

[1]*State Key Laboratory of Information Photonics and Optical Communications, Beijing University of Posts and Telecommunications, Beijing 100876, China*

[2]*Shandong Provincial Key Laboratory of Optical Communication Science and Technology, School of Physical Science and Information Technology, Liaocheng University, Liaocheng 252000, China*

[3]*Poly-Grames Research Center, Department of Electrical Engineering, École Polytechnique de Montréal, Montréal, QC H3T 1J4, Canada*

*E-mail: yqhuang@bupt.edu.cn

A novel uni-traveling-carrier photodetector (UTC-PD) structure with an integrated focusing-reflection (FR) mirror realized by a non-periodic concentric circular high-contrast grating (NP-CC-HCG), referred to as FR-UTC-PD, is proposed to enhance responsivity in conventional UTC-PDs. The FR-UTC-PD allows improving the responsivity by 36.5% at a 1.55-μm wavelength as compared to a UTC-PD without integrated an FR mirror with 84.59% reflectivity. For 40-μm-diameter PDs, the obtained 3-dB bandwidths are unaltered with values of 18 GHz at -3.0 V bias voltage. The radio-frequency (RF) output power and photocurrent are -1.77 dBm and 17.56 mA, respectively, at 10 GHz and the -6.0 V bias voltage.

The high-speed and high-responsivity photodetectors operating at long wavelength regimes have been widely used in wireless communication system, broadband optical communication system, and high-frequency measurement systems.[1-2] The uni-traveling-carrier photodetector (UTC-PD)[3-4] only utilizes electrons as active carriers and thus easily exhibits high-speed and high-output characteristics simultaneously, which has been of great interest both for research[5-8] and broad ranges of applications.[9-14] Moreover, high-speed UTC-PDs also need high-responsivity to prevent the waveform distortion of the high-speed signals and thus degrade the total power consumption of the system. Therefore, it is very important to adopt methods to further improve the bandwidth and responsivity performance for UTC-PDs. For instance, a UTC-PD with the downscaling absorber thickness and absorption area starting with 86 nm, 13 μm$^2$ scaled down to 30 nm, 5 μm$^2$, achieves 3-dB bandwidths from 235 GHz to 310 GHz while having the responsivities from 0.126 A/W to 0.07 A/W, respectively, at a 1.55-μm wavelength.[5-6] In order to achieve higher responsivity, a UTC-PD with a 1.2-μm-thick InGaAs absorber exhibits the responsivity up to 1.0 A/W, with a 3-dB bandwidth of only 9 GHz under low photocurrent.[15] Hence, balancing the trade-off between the bandwidth and responsivity in UTC-PDs has become an active research area.

Recently, various structures have been explored to address the bandwidth-responsivity trade-off problem in UTC-PDs. One is the waveguide structure UTC-PDs (WG-UTC-PDs). The effective absorption length for incident light is realized by increasing the WG length, so the responsivity is improved without sacrificing the device bandwidth. Another candidate is monolithic or quasi-monolithic integrated vertical dual-mesa (VDM) structure UTC-PDs by using different mirrors to increase the light length passing through the absorber. Figure 1 shows the relationship between responsivity and absorber thickness of several related works.[10,16-26] As it can be seen that the WG-UTC-PDs, which are near-ballistic UTC-PD (NBUTC-PD),[16] evanescent WG-UTC-PD,[17] WG-UTC-PD on silicon-on-insulator (SOI) substrate,[18] WG-UTC-PD on silicon-on-diamond (SOD) substrate[19] and WG-UTC-PD integrated on SOI nano-waveguide,[20] demonstrate many responsivity advantages in comparison with UTC-PDs with VDM structures. Compared to UTC-PDs with waveguide structures, some VDM structures also exhibit the improvement of the responsivity characteristic, such as resonant-cavity-enhanced UTC-PD (RCE-UTC-PD),[21] hexagonally shaped dual-mesa UTC-PD that integrated a total-reflection mirror (TR-UTC-PD),[22] back-to-back UTC-PD (B-to-B UTC-PD),[23] charge-compensation modified UTC-PD (CC-MUTC-PD)[9-10,24-25] and high-reflectivity UTC-PD (HR-UTC-PD).[26] A relatively high responsivity is realized by those different kinds of structures without sacrificing the bandwidth performance. However, many problems arise, such as the epitaxial growth time and cost for a distributed Bragg reflector (DBR) mirrors, the fabricated complexities for a monolithic integrated device and even the coupling problem during the measurement for a waveguide-integrated device.

In addition, the periodic strip pattern high-contrast grating (HCG) is reported in place of a conventional DBR to demonstrate an equivalent reflectivity.[27-28] Compared to a DBR mirror, achieving equivalent reflectivity by periodic strip HCG can potentially reduce the epitaxial layer thickness and simplify the subsequent fabrication procedure. So far, the periodic strip HCG has been extensively integrated with optoelectronic devices, such as vertical-cavity surface-emitting lasers (VCSELs), tunable VCSELs, tunable filters, high-Q optical resonators and low-loss hollow-core waveguides.[28-30] Furthermore, non-periodic HCG with strip patterns, blocky patterns, cylindrical patterns and spherical patterns can also exhibit high reflectivity, beam steering ability, focusing ability as well as optoelectronic integration.[31-33] In addition, another type of periodic or non-periodic concentric-circular HCGs not only provide high reflectivity, but also achieve distinct point focusing ability, while readily realizing wafer-scale integration of Si-based optoelectronic devices.[34-36]

In this letter, a high focusing-reflection (FR) uni-traveling-carrier photodetector (FR-UTC-PD) structure is proposed by using wafer-bonding technology to improve the responsivity of UTC-PDs without sacrificing the bandwidth performance. The schematic diagram of the FR-UTC-PD is shown in Fig. 2 (a). As it can be seen that the UTC-PD structure is integrated with a non-periodic concentric circular high-contrast grating (NP-CC-HCG) mirror on SOI substrate. The high-speed and high responsivity properties of the FR-UTC-PD are achieved simultaneously by the UTC-PD and the high FR NP-CC-HCG mirror, respectively. Hence, we can design a thinner InGaAs absorber to reduce the carrier transit time, while downscaling the absorption area to yield a lower resistance-capacitance (RC) time, so with the combination of those two aspects a higher bandwidth will be realized. In addition, the lower responsivity in UTC-PD caused by the aforementioned factors can be replenished by the bottom NP-CC-HCG mirror on SOI substrate.

The NP-CC-HCG mirror and its FR ability configuration are shown in Figs. 2 (a) and (b), respectively. It can be seen that the NP-CC-HCG mirror is composed of a set of non-periodic concentric circular Si grating units surrounded by air and the bottom adjacent $SiO_2$ layer. The 500-nm-thick Si grating ($t_g$) with a high refractive index of 3.48 serves as high-index layer, while the air and 500-nm-thick $SiO_2$ with the refractive index of 1 and 1.46 represent the low-index cladding layers. The reflected intensity of the NP-CC-HCG mirror depends on the difference between the high and low-index layers, i.e. the larger the indices difference is, the larger reflected intensity becomes. Moreover, the grating thickness ($t_g$), period ($\Lambda$) and duty cycle ($\eta$, defined as the grating width divided by the grating period) together determine the optical properties (such as phase and reflectivity) of the NP-CC-HCG mirror. However, due to the fixed grating thickness in our design, thus the grating period and duty cycle are the primary factors for optical properties. Meanwhile, unlike the periodic CC-HCG mirror realizing its optical properties by using the fixed grating period and duty cycle, the NP-CC-HCG mirror exhibits those properties by means of wave front control, i.e. obtaining the concentric circular grating aperiodic by adaptively modifying the local period and duty cycle. Therefore, the reflected light of an NP-CC-HCG mirror will realize a focusing spot when the total phase distribution satisfies Eq. (1):[30-31]

$$\phi(x) = \frac{2\pi}{\lambda}\left(f + \frac{\phi_{max}}{2\pi}\lambda - \sqrt{x^2 + f^2}\right) \quad (1)$$

where $f$ is the focal length, $\lambda$ is the wavelength, $x$ is the radius direction of an NP-CC-HCG mirror, and $\phi_{max}$ is the maximum phase change (the phase difference between the center and the very edge of an NP-CC-HCG).[30]

Along $x$ direction, the ideal phase distribution of the NP-CC-HCG mirror is calculated by Eq. (1) and the actual phase distribution of each designed discrete concentric circular grating bar is simulated, the obtained results are shown in Fig. 3 (a). Obviously, the discrete grating bars along the $x$ direction coincides exactly with the theoretical calculation results. Furthermore, the FR characteristics of the NP-CC-HCG mirror are modeled and simulated using COMSOL that is based on the finite element method (FEM) [37] with a preset focal length of 12 μm[34] to avoid the duplicate calculation of the mass data, and the relevant result is shown in Fig. 3 (b). It is clear that the focus spot along the $z$-axis is 11.65 μm which agrees reasonably with the 12-μm preset value. Besides, the full-width-half-maximum (FWHM) of the field distribution at the reflected focus plane is 0.8701 μm, while the FR efficiency at the reflected plane is 92.1%.

To further study the FR ability experimentally, an NP-CC-HCG mirror with 500-μm diameter and 400-μm focal length is fabricated and the scanning electron microscope (SEM) images in various locations are shown in Figs. 4 (a) and (b), respectively. Moreover, Fig. 4 (c) demonstrates the FR ability of the NP-CC-HCG mirror at a 1.55-μm wavelength radially polarized light and a 3.56-mW input power, which is represented by the power distribution at the reflected focal plane. It is clear that the reflected optical power approximately presents Gaussian distribution along the $x$ direction and the peak value locates at the middle position of the NP-CC-HCG mirror, which reveals that this mirror has an excellent focusing ability.

The maximum reflected optical power at the focal spot is 463.5 µW and the FWHM is ~140 µm, while the FR efficiency of 84.59% is achieved by integration.

The UTC-PD structure mainly consists of a 640-nm-thick p-type InGaAs absorber, a 200-nm-thick depleted InGaAs absorber and a 500-nm-thick n-type InP collector. The VDM structure is achieved by conventional photolithography and wet-etching technologies. The ground-signal-ground (GSG) coplanar waveguide (CPW) electrodes are landed on the polyimide passivation layer for high-frequency measurement. The realization of the FR-UTC-PD is based on micron-level-thick benzocyclobutene (BCB) material. However, due to the 400-µm focal length limitation of the NP-CC-HCG mirror, the removing thickness of the InP-substrate and BCB bonding-thickness are carefully considered before wafer-bonding process. It is worth noticing that the BCB bonding layer should be equal to or greater than 500 nm to prevent the disturbance caused by the non-uniform thickness of the circular grating from affecting the focusing position of the reflected light. The whole bonding process is in nitrogen ambient and then the FR-UTC-PD is picked out when the temperature of the bonding cavity degrades to room temperature. Thereafter, the measurement is conducted at room temperature.

Figure 5 (a) shows the measured output photocurrent versus the input power for the UTC-PD and the FR-UTC-PD at the bias voltage of -3.0 V and a 1.55-µm wavelength excitation. It can be seen that the output photocurrent increases monotonously with increasing the input power up to a saturation point, while the linear parts reveals the responsivity properties of the two. The responsivity is 0.63 A/W and 0.86 A/W for UTC-PD and FR-UTC-PD, respectively. Compared to the UTC-PD without the NP-CC-HCG mirror, the responsivity of FR-UTC-PD is increased by 36.5%. This is because the NP-CC-HCG mirror makes the FR-UTC-PD get almost twice the effective absorption efficiency from the incident light than that of UTC-PD. According to Eq. (2)[38], the actual reflectivity of the NP-CC-HCG mirror bonded in FR-UTC-PD is 63% based on the measured responsivity of the UTC-PD and FR-UTC-PD.

$$r = (R/R_0 - 1)e^{\alpha d} \quad (2)$$

where $R_0, R$ are the responsivity of UTC-PD and FR-UTC-PD, $\alpha, d$ are the absorption coefficient of $In_{0.53}Ga_{0.47}As$ layer and the thickness of the active layer, $r$ is the reflectivity of the NP-CC-HCG mirror. It is obvious that there is 21.59% illumination loss, which might be incurred by the bonding and the measurement process, comparing with the measured value (84.59%) of the NP-CC-HCG mirror before bonding. Therefore, the responsivity of the FR-UTC-PD can also be further improved by reasonably designing the focal length of the NP-CC-HCG mirror or precisely controlling the BCB bonding thickness together with the removing thickness of the InP substrate, which makes the reflected light to be absorbed totally by the absorber of the UTC-PD. In addition, depositing single layer $SiO_2$ or $SiN_x$, or double $SiO_2/SiN_x$ antireflection coating on the top side (i.e. device surface) to reduce the optical reflection loss are other methods to improve the responsivity.

The measured RF output power and the photocurrent at 10 GHz for an FR-UTC-PD with a 40-μm diameter at various bias voltages and a 1.55-μm wavelength is shown in Fig. 5 (b). The obtained output RF powers are -6.97 dBm, -3.98 dBm, -2.75 dBm and -1.77 dBm corresponding to the photocurrents of 13.3 mA, 16.37 mA, 17.35 mA and 17.56 mA at the bias voltages of -3.0 V, -4.0 V, -5.0 V and -6.0 V, respectively. It should be noted that the photocurrent increases monotonously with increasing the bias voltage, while the output RF power also increases and will gradually approach the ideal RF power if there is a thermal-conductivity submount. However, the RF output power and photocurrent are relatively low compared to the ideal value due to the following reasons.

The first and primary reason is the Joule heat problem during the measurement process. Because there is not any high-thermal-conductivity submount in our device, so the device generates lots of Joule heat in the junction with increasing the bias voltages during the measurement, which makes the output power and the photocurrent readily come up to saturation in advance and even causes the device to fail. The second reason may come from the BCB bonding process. Due to the influence of mechanical vibration in the process of grinding and polishing the InP substrate, together with the uniaxial pressure resulting from the self-fixture during BCB bonding process, the Ohmic contact resistances become larger to result in serious Joule heat problem. The third reason might be the imperfect design of the intrinsic epitaxial structure and the human factors incurred during the fabricated process for the UTC-PD. Therefore, we can optimize the device epitaxial structure in the subsequent design by adding a cliff layer[39] with a relative higher n-type doping (~$10^{18}$ cm$^{-3}$ order of magnitude) between the spacer and the collector to reduce the deleterious space charge effect at high photocurrent levels. For the etching process during device fabrication, we can use inductively coupled plasma etching (ICP) in place of wet-etching to obtain the smooth sidewall and then to degrade the effect of sidewall leakage current and dark current.

In a word, in order to improve the RF output power, it is quite necessary to adopt flip-chip techniques (such as bonding FR-UTC-PD on aluminum nitride (AlN) or diamond surmounts) to reduce the thermal dissipation[40,10] at high bias voltages. In addition, optimizing device structure to increase the output photocurrent levels, reducing the effect of artificial factors in the process of fabrication and measurement, and adopting modulation depth enhancement technique[10] in place of the conventional heterodyne measuring method will further expand the RF output power.

The measured frequency responses for the FR-UTC-PD and UTC-PD with a 40-μm diameter in a series of low output photocurrent levels are shown in Fig. 5 (c). It can be seen that the 3-dB bandwidths of all are 18 GHz at a bias voltage of -3.0 V and a 1.55-μm wavelength source when the photocurrents increase from 10 μA to 190 μA. The unaltered bandwidths indicate that there is no effect on the bandwidth for a UTC-PD with or without an NP-CC-HCG mirror. Furthermore, the no apparent saturation states indicate that the PD has a good operational state under low photocurrent levels. In theory, the 3-dB bandwidth is mainly determined by the carrier transit time and the RC limited time. Besides, the carrier transit time limited bandwidth is a constant value in our PD due to the constant thickness of the absorber. Therefore, a good choice to improve the bandwidth is to design a PD with a smaller diameter. Because a smaller diameter corresponds to a smaller junction capacitance, which leads to a larger RC limited bandwidth. Besides, another method is to apply a higher bias voltage, which will extend the depletion region to the absorber and thus shorten the carrier transit time in the absorption region. Whereas, it is noticed that the excessive bias voltage will affect the reliability of the device and even make it fail.

In conclusion, an FR-UTC-PD via integrating an NP-CC-HCG mirror with a UTC-PD structure is reported to overcome the responsivity-bandwidth trade-off in conventional UTC-PDs. The FR ability of the NP-CC-HCG mirror is analyzed by simulation and experiment. The simulated NP-CC-HCG mirror achieves an 11.65-μm focal length, while the FWHM is 0.8701 μm and the FR efficiency is 92.1%. Further, the fabricated NP-CC-HCG mirror with a 400-μm focal length demonstrates an FWHM of ~140 μm simultaneously obtaining the corresponding FR efficiency of 84.59%, and has 21.59% illumination loss after bonded in FR-UTC-PD. For the integrated FR-UTC-PD, the responsivity is 0.86 A/W showing an improvement of 36.5% at a 1.55-μm wavelength. The 3-dB bandwidth for a 40-μm-diameter FR-UTC-PD is an unaltered value of 18 GHz at -3.0 V bias voltage while the output RF power is -1.77 dBm with a photocurrent of 17.56 mA at 10 GHz under a -6.0 V bias voltage. This design fully manifests the advantages of designing the devices independently and paves the way for high-performance CMOS-compatible integrated optoelectronic devices and will play a role in improving the responsivity for the other types of III-V and IV near-infrared and mid-infrared photodetectors[41] in future.

### Acknowledgments

This work is supported by the National Natural Science Foundation of China (NSFC) under the Grant Nos. 61574019, 61674018 and 61674020. The authors thank Jiarui Fei and Tao Liu for the assistance with measurements and fruitful discussions.

Figure Captions

**Fig. 1.** Summary of the responsivity versus absorber thickness for different kinds of UTC-PD structures. (WG: waveguide; VDM: vertical dual-mesa.)

**Fig. 2.** (a) Schematic cross-sectional layout of the FR-UTC-PD, consisting of an SOI-based NP-CC-HCG mirror and a top UTC-PD epitaxial structure. (b) The typical NP-CC-HCG mirror structure located in the white dashed box of Fig. 2 (a) and the FR behavior realized by wave front control.

**Fig. 3.** (a) The calculated (the red line) and simulated (the blue solid circles) results of the phase distribution and (b) the FR ability simulation with a 12-µm preset focal length for an NP-CC-HCG mirror.

**Fig. 4.** The SEM images at (a) the center position and (b) the non-periodic position for the NP-CC-HCG mirror. (c) The FR ability of the NP-CC-HCG mirror represented by the power distribution.

**Fig. 5.** (a) The output photocurrent versus input power for UTC-PD and FR-UTC-PD. Inset: the optical micrograph of a fabricated PD. (b) The RF output power versus photocurrent at 10 GHz for the FR-UTC-PD with a 40-µm diameter at different bias voltages. The black dash line is the parallel line of the ideal RF power indicated a linear photocurrent-power relation. (c) The frequency response for UTC-PD and FR-UTC-PD with the same diameters at various output photocurrent levels.

**Fig. 1.**

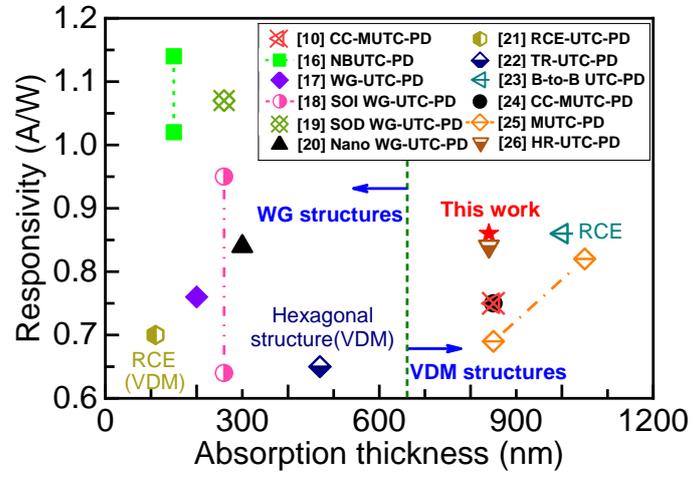

**Fig. 2.**

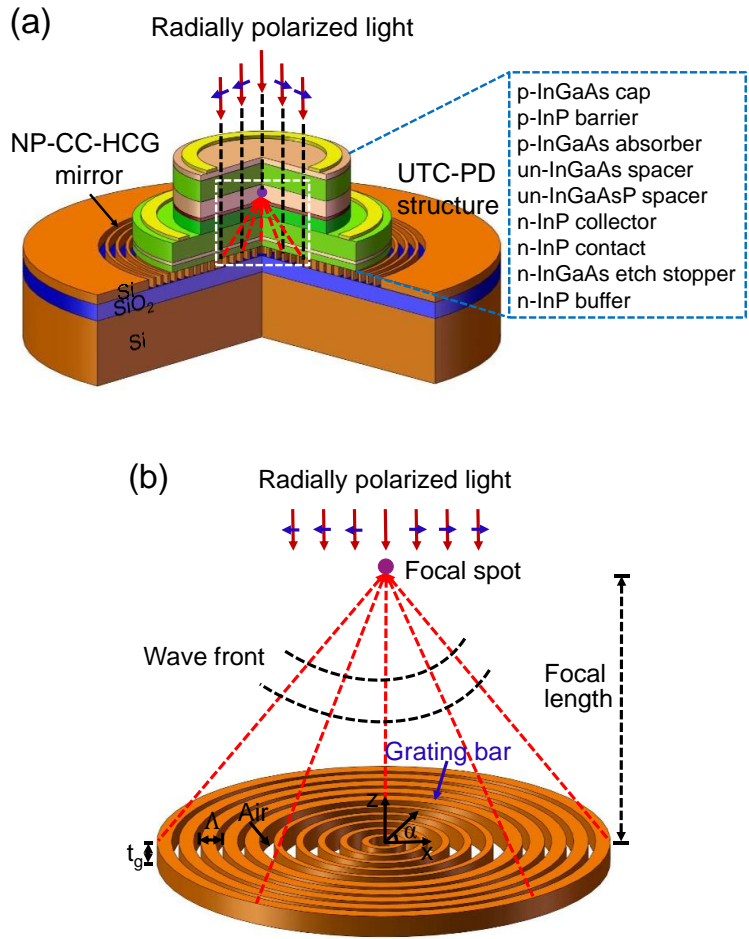

**Fig. 3.**

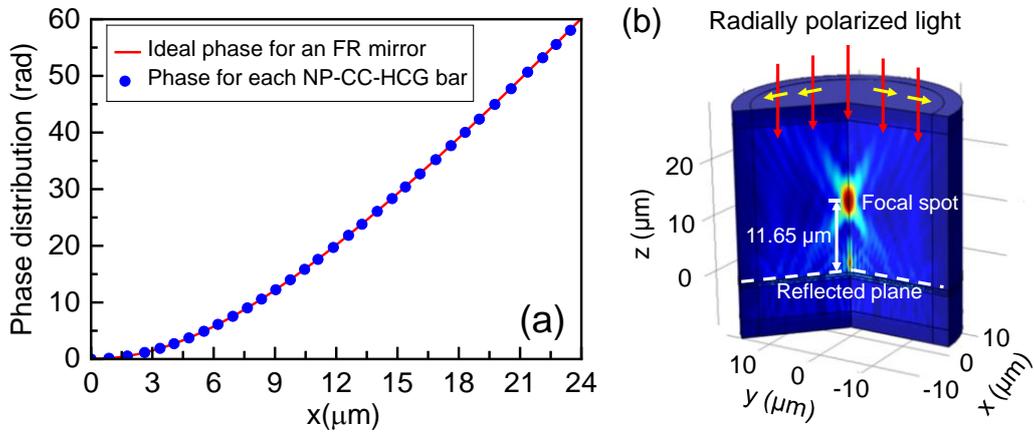

**Fig. 4.**

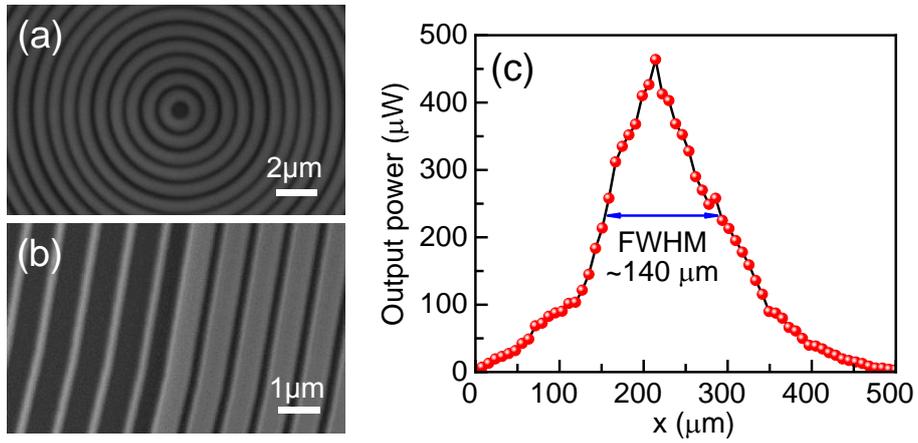

**Fig. 5.**

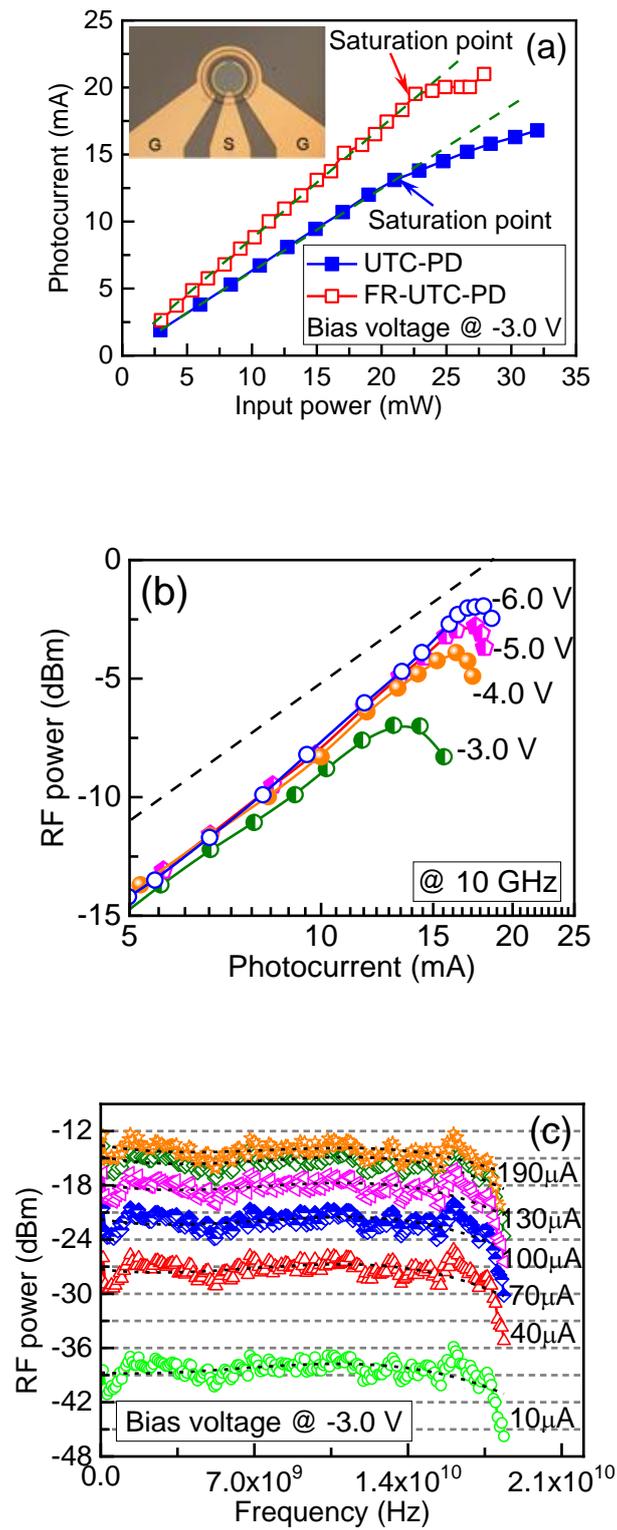